\documentclass[aps,prl,twocolumn]{revtex4}

\usepackage{graphicx,amsmath}

\begin{document}

\newcommand{\be}{\begin{equation}}
\newcommand{\ee}{\end{equation}}
\newcommand{\bea}{\begin{eqnarray}}
\newcommand{\eea}{\end{eqnarray}}
\newcommand{\qea}{q_{\scriptscriptstyle{\rm EA}}}
\newcommand{\mc}{\mathcal}
\newcommand{\eps}{\varepsilon}
\newcommand{\s}{\sigma}
\newcommand{\ds}{\Delta\sigma}
\newcommand{\hw}{h^{\rm W}}
\newcommand{\sw}{\sigma^{\rm W}}
\newcommand{\<}{\langle}
\renewcommand{\>}{\rangle}

\title{Measuring the fluctuation-dissipation ratio in glassy systems
with no perturbing field}

\author{F. Ricci-Tersenghi}

\affiliation{Dipartimento di Fisica, SMC and INFM, Universit\`a di
Roma ``La Sapienza'', P.le Aldo Moro 2, 00185 Roma, Italy}

\date{\today}

\begin{abstract}
A method is presented for measuring the integrated response in Ising
spin system without applying any perturbing field.  Large-scale
simulations are performed in order to show how the method works.  Very
precise measurements of the fluctuation-dissipation ratio are presented
for 3 different Ising models: the 2-dimensional ferromagnetic model,
the mean-field diluted 3-spin model, and the 3-dimensional
Edwards-Anderson model.
\end{abstract}

%\pacs{75.50.Lk}

\maketitle

Disordered and frustrated models are a fascinating but still poorly
understood subject in contemporary statistical mechanics.  The
interest in these systems also comes from their many interdisciplinary
applications: from the physics of glass-former liquids to that of
polymers and biomolecules, from the description of error correcting
codes to the study of the computational complexity and phase
transitions in theoretical computer science.

Here we will use the term {\em glassy system} for a generic model
showing very slow relaxation to equilibrium \cite{REVIEW}.  Because of
the huge equilibration time, a glassy system may be in the out of
equilibrium regime for all the experimental times.  Then a complete
understanding of this regime is what one needs in order to correctly
describe a real slow-evolving material.  Moreover, numerical studies
of the off-equilibrium regime do not suffer from finite-size effect
since very large sizes can be used.  They present finite time
corrections which can be usually kept under control, thus allowing for
better numerical estimations.

Among the numerical methods that can be used in the out of equilibrium
regime, the study of the so-called {\em fluctuation-dissipation ratio}
(FDR) \cite{CUKU} has been shown to be a very successful one
\cite{MPRR98,REVIEW_FDR}.  This method is based on the comparison of
how spontaneous and induced fluctuations relax.  Actually one measures
an autocorrelation function $C(t,s)$~\footnote{We always consider
correlation functions connected, i.e.\ $\lim_{t\to\infty}
C(t,s)\!=\!0\quad\forall s<\infty$, and normalized, $C(t,t)\!=\!1$.}
and the associated response function $R(t,s)$ and defines the FDR
$X(t,s)$ through the formula
\be
T\,R(t,s) = X(t,s)\;\partial_s C(t,s) \quad ,
\ee
where $T$ is the temperature.  At equilibrium the
fluctuation dissipation theorem (FDT) holds, implying $X=1$.

In the large times limit --- $s,t\!\to\!\infty$ with $C(t,s)\!\to\!q$
--- the FDR $X(t,s)$ converges to the limiting function $X(q)$.  The
physical meaning of the function $X(q)$ has been explained in
Refs.~\cite{FMPP}, where it has been shown that under some hypothesis
(stochastic stability) the equation
\be
X(q) = x(q) \equiv \int_0^q P(q') dq'
\label{eq:Xx}
\ee
holds.  In Eq.(\ref{eq:Xx}) $P(q)$ represents the overlap pdf in the
threshold states, that is the states reached by the out of equilibrium
dynamics on very large times, which could be different from the
thermodynamical state \cite{REVIEW}.  It has been conjectures that the
effective temperature $T_{\rm eff} = T/X$ plays a central role in
off-equilibrium glassy systems \cite{CUKUPE}.

In numerical simulations the punctual response function $R(t,s)$ is
very noisy, while a much better signal can be obtained for the
integrated response function
\be
\chi(t,t_w) = T \int_{t_w}^t R(t,s)\,ds \quad .
\label{eq:chi1}
\ee
With respect to the usual definition, the temperature $T$ has been
added in the above equation in order to simplify the notation in the
following formulae and to have a well defined $\chi(t,t_w)$ in the $T
\to 0$ limit.  In the large time limit, Eq.(\ref{eq:chi1}) can be
rewritten as
\be
\chi(C) = \int_C^1 X(q)\,dq \quad .
\ee
So the FDR can be simply written as $X(C)=-\partial_C \chi(C)$.

The aim of this Letter is to propose and to show the efficacy of a
very precise method for measuring the integrated response $\chi(C)$
and the FDR $X(C)$ in spin models.

Up to now the best protocol for measuring $\chi(C)$ in spin systems
has been the following \cite{MPRR98,BARRAT}:
\begin{itemize}
\item initialize the system in a random configuration;
\item quench the system at a temperature $T<T_c$ and evolve it for
  $t_w$ Monte Carlo sweeps (MCS);
\item switch on a random magnetic field of very small intensity $h$
  and continue evolving the system while measuring $C_h(t,t_w)$ and
  $\chi_h(t,t_w)$.
\end{itemize}
The parametric plot of $\chi_h(t,t_w)$ versus $C_h(t,t_w)$ converges
to the function $\chi(C)$ in the limit $t_w\!\to\!\infty$ and
$h\!\to\!0$.  Even when extrapolations can be safely done, they always
require a large numerical effort: for example, in order to correctly
take the $h \to 0$ limit, the whole simulation must be repeated for
many $h$ values in the linear response regime.  Moreover in frustrated
systems like spin glasses the response may have strong non-linearities
even for very small probing fields and it is usually very hard to
predict \textit{a priori} which is the linear response regime.
Furthermore in out of equilibrium simulations the size of the linear
response regime may change with the age of the system: A fair
conjecture is that it decreases for larger times.  If this would be
true, extrapolations to the interesting limit would become still more
difficult.

For all these reasons we consider of primary importance the
development of a method which allows one to calculate the linear
response in a spin system without applying any probing field.  After
having taken {\em analytically} the $h \to 0$ limit, one is left only
with the $t_w \to \infty$ limit.  This limit will be somehow
unavoidable as long as the only way for aging a glassy system will be
to simulate it for a long time~\footnote{Accelerated dynamics can be
used only in some particularly simple cases}.

Inspired by a recent work by Chatelain \cite{CHAT}, we write down an
analytical expressions giving the integrated response $\chi(t,t_w)$ in
a simulation with no probing field~\footnote{In the first version of
Ref.~\cite{CHAT}, the only available at the time this work was
finished, only the punctual response function $R(t,s)$ was studied and
the calculations had some errors, making the final result
unreliable.}.

Let us specialize on systems with $N$ Ising spins and Hamiltonian
$\mc{H}_0$ (generalization to Potts variables is straightforward
\cite{FDT_POTTS}).  The Hamiltonian $\mc{H}_0$ may contain some
quenched disorder, but we do not need to specify it, since our
calculations hold for a generic $\mc{H}_0$, either disorder or not
disordered.  In the former case the final result can be eventually
averaged over the quenched disorder distribution, but following
formulae are valid for any given disorder realization.

When the probing field is switch on the Hamiltonian becomes $\mc{H} =
\mc{H}_0 - \sum_{i=1}^N h_i \s_i$, where $h_i$ are i.i.d.\ random
variables with $\overline{h_i}=0$ and $\overline{h_i h_j} =
h^2\,\delta_{i,j}$.  For simplicity we define $h_i = h\,\eps_i$ with
$\overline{\eps_i}=0$ and $\overline{\eps_i\,\eps_j} = \delta_{i,j}$.

The FDR for the observable $A(t) = \sum_i \eps_i \s_i(t)$ is given in
terms of the correlation and response functions
\bea
N C(t,s) &=& \overline{\<A(t) A(s)\>} = \sum_i \<\s_i(t)\s_i(s)\>\;,\\
N R(t,s) &=& \overline{\frac{\partial\<A(t)\>}{\partial h(s)}} =
\sum_i \overline{\eps_i \sum_j \frac{\partial\<\s_i(t)\>}{\partial
h_j(s)} \frac{\partial h_j}{\partial h}} = \nonumber\\
 &=& \sum_{i,j} \overline{\eps_i \eps_j}
\frac{\partial\<\s_i(t)\>}{\partial h_j(s)} = \sum_i
\frac{\partial\<\s_i(t)\>}{\partial h_i(s)}\;,
\eea
where $\<\,\cdot\,\>$ represents the average over thermal histories.

We use a discrete-time dynamics as the one taking place in a Monte
Carlo simulation.  The time $t$ counts the number of single spin
updates and not the number of Monte Carlo sweeps (which is then
$t/N$).  The function $I(t)$ gives the index of the spin to be updated
at time $t$, and so it depends on the updating rule (e.g.\ random or
sequential).  At the $t$-th time step the spin $\s_i$ with $i=I(t)$ is
updated according to heat-bath probabilities
\be
\text{prob}(\s_i=\s) = \frac{\exp[\beta \s (\hw_i+h_i)]}
{2 \cosh[\beta (\hw_i+h_i)]}\quad,
\ee
where $\beta$ is the inverse temperature and the Weiss field $\hw_i$
takes into account the effect of Hamiltonian $\mc{H}_0$ on the spin to
be updated.  For example, in the case of 2-spin interacting
Hamiltonians the Weiss field is given by $\hw_i = \sum_{j \neq i}
J_{ij} \s_j$.

Under this dynamics the expectation value of the $j$-th spin a time
$t$ is given by
\be
\<\s_j(t)\> = {\rm Tr}_{\vec\s(t')}\left[ \s_j(t)
\prod_{t'=1}^t W_{I(t')}\Big( \vec\s(t') | \vec\s(t'\!-\!1) \Big)
\right] \label{eq1}
\ee
where $\vec\s$ is a short-hand notation for the $N$ spins
configuration, the trace is over all the trajectories $\vec\s(t')$
with $1 \le t' \le t$, and the transition probability is given by
\be
W_i(\vec\s | \vec\tau) = \frac{\exp[\beta \s_i (\hw_i+h_i)]}
{2 \cosh[\beta (\hw_i+h_i)]} \prod_{j \neq i} \delta_{\s_j,\tau_j}\ .
\ee
Note that $\hw_i(\vec\s) = \hw_i(\vec\tau)$ since it does not depend
on the spin in $i$.  The transition probability $W_i$ only depends on
the perturbing field on site $i$, such that
\be
\left. \frac{\partial W_i(\vec\s | \vec\tau)}{\partial h_j}
\right|_{h=0} = \delta_{i,j}\; W_i(\vec\s | \vec\tau)\,\beta\,\left(\s_i
- \sw_i\right) \ ,
\ee
where we have defined $\sw_i \equiv \tanh(\beta \hw_i)$.

Now we suppose that an infinitesimal probing field $h_k$ on site $k$
is switch on at time $t_w$: $h_k(t) = h\,\theta(t-t_w)$.  This means
that the transition probability $W_k$ (and only this one) will depend
on the perturbing field for all times larger than $t_w$.
Differentiation of Eq.(\ref{eq1}) with respect to this field yields
the integrated response
\begin{widetext}
\bea
\chi_{jk}(t,t_w) &=& T \left.\frac{\partial \<\s_j(t)\>}{\partial h}
\right|_{h=0} = {\rm Tr}_{\vec\s(t')} \left[ \s_j(t) \prod_{t'=1}^t
W_{I(t')} \Big(\vec\s(t') | \vec\s(t'\!-\!1)\Big) \sum_{s=t_w+1}^t
\delta_{I(s),k}\,\left(\s_k(s) - \sw_k(s)\right) \right] = \nonumber\\
&=& \<\s_j(t)\,\ds_k(t,t_w)\> \qquad \text{ with } \qquad \ds_k(t,t_w)
= \sum_{s=t_w+1}^t \delta_{I(s),k} \left(\s_k(s) - \sw_k(s) \right)\;.
\label{resp}
\eea
\end{widetext}
The correlation in Eq.(\ref{resp}) is what one has to measure in a
numerical simulation with no perturbing field in order to get the
integrated linear response.

Few comments are in order.  The time-integrated quantity $\ds_k$ only
gets contributions when the spin $\s_k$ is updated, so most of the
times it is unchanged.  The contributions summed up in $\ds_k$ are the
differences among the actual value of the spin $\s_k$ and the expected
one $\sw_k$.  So $\ds_k$ is a random variable with zero mean,
$\<\s_k\>=\sw_k \Rightarrow \<\ds_k\>=0$, and variance $\<\ds_k^2\>
\propto (t-t_w)$.

In the $T=0$ limit Eq.(\ref{resp}) has a nice and simple physical
interpretation.  Since for $T=0$ we have that $\s_k - \sw_k = \s_k
\delta_{\hw_k,0}$, then $\ds_k$ takes a contribution only when the
spin $\s_k$ has a zero Weiss field on it, i.e.\ it is free to respond
to an infinitesimal field.  If the Weiss field is different from zero
the spin is completely frozen and it can not respond to an
infinitesimal perturbing field.  So the integrated response in
Eq.(\ref{resp}) can be rewritten as a simple sum of correlation
functions $\chi_{jk}(t,t_w) = \sum_s^\prime \<\s_j(t) \s_k(s)\>$,
where the primed sum is over all the times larger then $t_w$ when
$\s_k$ is updated under a zero Weiss field, i.e.\ being a free spin.

We now present numerical results for 3 spin models which are believed
to belong to 3 different classes: ferromagnetic model in 2D
(coarsening system), diluted long-range 3-spin model with fixed
connectivity 4 (discontinuous spin glass) and Edwards-Anderson model
in 3D (continuous spin glass).  For each model we have checked that
the $\chi_h(C_h)$ curve measured with the perturbing field converges
for $h\!\to\!0$ to the one measured with the new method.  Hereafter
times will be counted in MCS.

\begin{figure}
\includegraphics[width=\columnwidth]{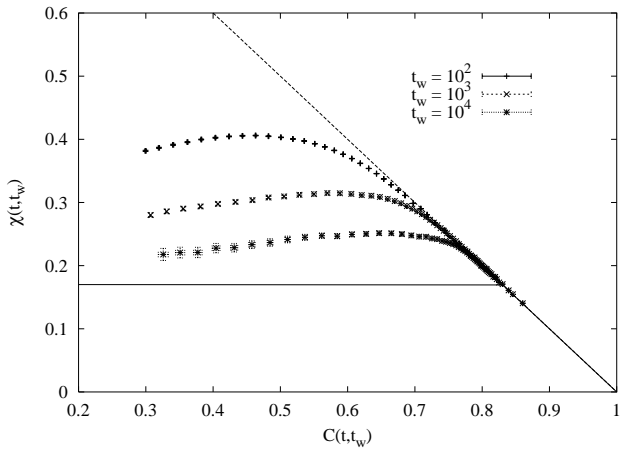}
\caption{FD plot for the 2D Ising ferromagnet at $T=2$. The horizontal
line is the analytic long time limit.}
\label{ferro}
\end{figure}

The first model is the ferromagnetic Ising model on the 2-dimensional
square lattice.  We have simulated at $T = 2 \simeq 0.88\, T_c$
systems of sizes $1000^2$ and $7000^2$ in order to check the absence
of any finite-size effect (the data we show are from the $1000^2$
samples).  For each waiting time, $t_w=10^2,10^3,10^4$, averages have
been taken over 100 different thermal histories, and the corresponding
$\chi(C)$ curves are shown in Fig.~\ref{ferro}.  The horizontal line
is the analytical prediction for the large times limit, $\chi =
1-m_{\rm eq}^2 = 0.17$.  Numerical curves are clearly compatible with
the analytical asymptote in the large times limit.

\begin{figure}
\includegraphics[width=\columnwidth]{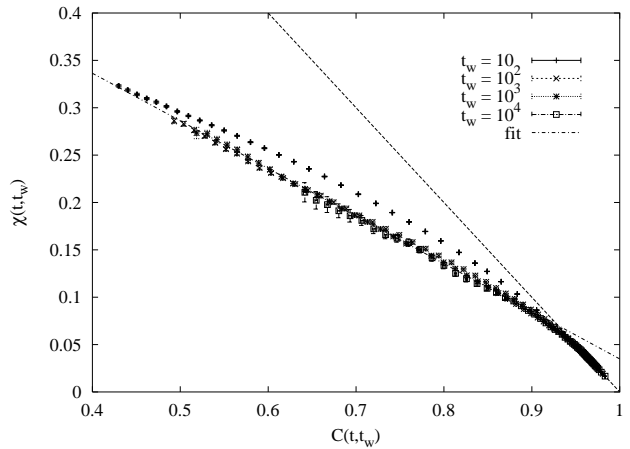}
\caption{FD plot for the long-range 3-spin model with fixed
  connectivity 4 at $T=0.5$.  The line $0.53745-0.50256\,C$ is the
  best linear fit to $t_w=10^3$ data with $C<0.9$.}
\label{3spin}
\end{figure}

The second model we studied is the 3-spin model defined on a random
hypergraph with fixed connectivity 4.  This model has been solved
analytically with a one-step replica symmetry breaking Ansatz in
Ref.~\cite{MORI_TFINITA}.  The dynamical critical temperature is
$T_d=0.755\pm0.01$.  We have run simulations for a size $N=999999$ at
temperature $T=0.5\simeq0.66\,T_d$ and the resulting $\chi(C)$ curve
is shown in Fig.~\ref{3spin}.  The number of samples used is 10 for
$t_w=10,10^2$, 50 for $t_w=10^3$ and 20 for $t_w=10^4$.  The straight
line in Fig.~\ref{3spin} is a linear fit to $t_w=10^3$ data in the
region $C(t,t_w) < 0.9$, which perfectly interpolates the data
($\chi^2/\text{d.o.f.}=0.82$).  It gives a Parisi breaking parameter
on threshold states equal to $m_{\rm th}(T=0.5) = 0.5 \pm 0.02$.  The
error is an estimate of systematic effects, mainly given by the slight
increase of $m$ with $t_w$.  Comparison of this value for $m_{\rm th}$
with corresponding static predictions will be done in
Ref.~\cite{MORI_TFINITA}.

\begin{figure}
\includegraphics[width=\columnwidth]{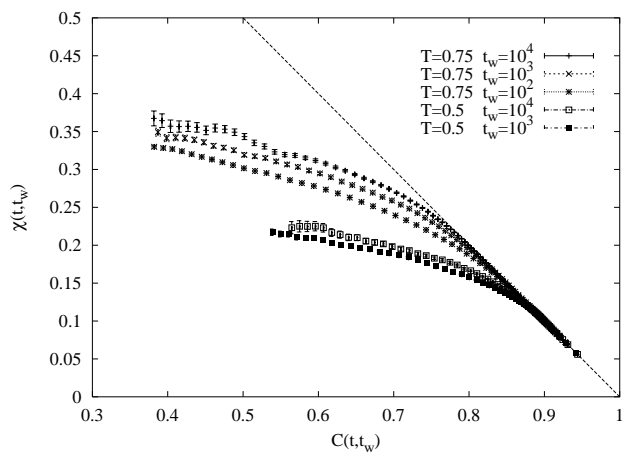}
\caption{FD plot for the 3D Edwards-Anderson model.}
\label{sg_3d}
\end{figure}

The third model we studied is the 3-dimensional Edwards-Anderson model
with $J=\pm1$ couplings, which undergoes a phase transition to a spin
glass phase at $T_c=1.14 \pm 0.01$~\cite{SUE}.  We have simulated
samples of size $L=20$ at temperatures $T = 0.75 \simeq 0.66\,T_c$ and
$T = 0.5 \simeq 0.44\,T_c$, for three different waiting times
$t_w=10^2,10^3,10^4$.  The results are shown in Fig.~\ref{sg_3d}.

\begin{figure}
\includegraphics[width=\columnwidth]{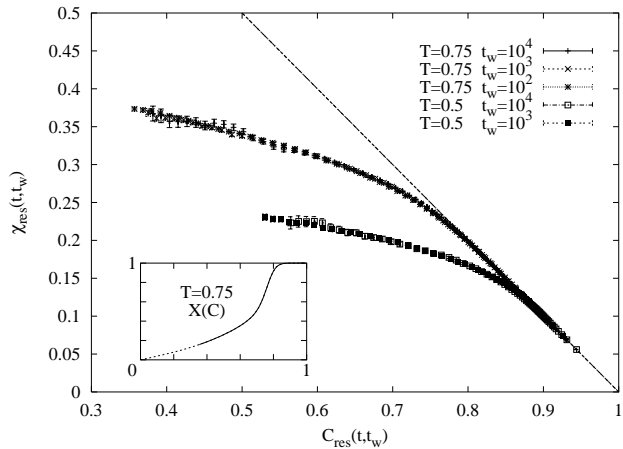}
\caption{Same as Fig.~\ref{sg_3d} with rescaled variables.  Inset: FDR
for $T=0.75$ obtained from the derivative of the rescaled data.}
\label{sg_3d_scal}
\end{figure}

For a given temperature the $\chi(C)$ curves look very similar in
shape, the main difference being the $t_w$-dependent Edwards-Anderson
order parameter $\qea(t_w)$, here defined as the point where the
$\chi(C)$ curve leaves the FDT line $1-C$.  In order to exploit all
the data we tried to collapse the curves before fitting.  The collapse
can be achieved either by shifting the curves such that the
$\qea(t_w)$ coincide, either by the following rescaling: $C_{\rm
res}(t,t_w) = \lambda\,C(t,t_w)/\qea(t_w), \chi_{\rm res}(t,t_w) =
1-\lambda\,(1-\chi(t,t_w))/\qea(t_w)$, with an arbitrary $\lambda$.
Both scaling are statistically acceptable.  In Fig.~\ref{sg_3d_scal}
we show the second one which is slightly better, with
$\lambda=\qea(10^4)$.

If the measured data are already in the asymptotic regime, i.e.\ the
scaling is valid for larger times, and since $\lim_{t_w \to \infty}
\qea(t_w) = \qea > 0$, we can conclude that the FDR is non trivial in
the 3-dimensional Edwards-Anderson model, with an $X(C)$ like the one
depicted in the inset of Fig.~\ref{sg_3d_scal} for $T=0.75$.

The Edwards-Anderson model is the one which took the great part of the
simulation time.  Indeed, in order to have reasonable error bars, we
ran at each of the two temperatures $10^4$ samples for $t_w \le 10^3$
and almost $3 \cdot 10^4$ samples for $t_w=10^4$.  Solely the
$t_w=10^4$ runs took the equivalent of more than 1 year of CPU-time on
a latest generation 2.0 GHz computer.  This is a consequence of the
fact that errors on the linear response $\chi(t,t_w)$ increases like
$\sqrt{t-t_w}$ and so the number of samples for keeping the error on
$\chi(C)$ constant increases more or less linearly with the waiting
time $t_w$.  We believe this is the main drawback of the present new
method for measuring the linear response: Although it is very
successful for small times, it becomes very noisy at larger times and
so it requires a huge statistics.

From this observation one could conclude that the usual old method of
measuring the response with a small perturbing field would eventually
remain the only valid one, but this is not true.  Very probably the
linear regime in the perturbing field $h$ decreases with the age of
the system.  In order to remain in the linear response regime one
should decreases the intensity of the perturbing field during the
simulation, thus increasing the error on the $\chi$ for late times.

Let us conclude with two remarks.  First remark: Having understood
that the integrated response can be written as a correlation function,
it should be clear that all the functions $C(t,t_w)$ and $\chi(t,t_w)$
can be calculated in the {\em same simulation} for any value of $t$
and $t_w$.  Moreover, being correlation functions self-averaging
quantities, it should be possible in principle to calculate them in a
{\em single simulation} of a sufficiently large sample.  Second
remark: The method presented here can be also used for any other Monte
Carlo simulation (e.g.\ glass-former particle systems).  The only
condition for using present analytical expressions is the discreteness
of time.

\textbf{Acknowledgments} Numerical simulations have been run on our
Linux cluster IDRA.  I thank ICTP for kind hospitality during the
completion of this manuscript.  Work supported in part by the European
Community's Human Potential Programme under contract
HPRN-CT-2002-00307, Dyglagemem.

\end{document}